\documentclass[twocolumn,trackchanges]{aastex701}

\usepackage[dvipsnames]{xcolor}
\usepackage{amsmath}
\usepackage[title]{appendix} 

\newcommand{\paa}{Pa{$\alpha$} }

\begin{document}


\title{Sequential Star-Formation and the Near-Infrared Stellar Populations within N159}

\author[0000-0002-2667-1676]{Nolan Habel}
\affiliation{Jet Propulsion Laboratory, California Institute of Technology, 4800 Oak Grove Dr., Pasadena, CA 91109, USA}
\affiliation{Department of Astronomy, California Institute of Technology, Pasadena, CA 91125; USA}
\email[show]{nhabel@caltech.edu}

\author[0000-0003-2954-7643]{E.\ Sabbi}
\affiliation{NSF NOIRLab: Tucson, Arizona, US}
\email{}

\author[0000-0002-6091-7924]{Peter\ Zeidler} 
\affiliation{AURA for the European Space Agency}
\affiliation{Space Telescope Science Institute, 3700 San Martin Drive, Baltimore, MD 21218, USA}
\email{}

\author[0000-0001-8538-2068]{Tuila Ziliotto} 
\affiliation{NSF NOIRLab: Tucson, Arizona, US}
\email{}

\author[0000-0002-1437-4463]{Laurie Chu} 
\affiliation{NSF NOIRLab: Tucson, Arizona, US}
\email{}

\author[0000-0002-0522-3743]{Margaret Meixner}
\affiliation{Jet Propulsion Laboratory, California Institute of Technology, 4800 Oak Grove Dr., Pasadena, CA 91109, USA}
\email{}

\author[0000-0001-7906-3829]{Guido De Marchi}
\affil{European Space Research and Technology Centre, Keplerlaan 1, 2200 AG Noordwijk, The Netherlands}
\email{gdemarchi@esa.int}

\author[0000-0003-4870-5547]{O.\ C.\ Jones}
\affiliation{UK Astronomy Technology Centre, Royal Observatory, Blackford Hill, Edinburgh, EH9 3HJ, UK}
\email{}

\author[0000-0002-2954-8622]{Alec S.\ Hirschauer}
\affil{Department of Physics \& Engineering Physics, Morgan State University, 1700 East Cold Spring Lane, Baltimore, MD 21251, USA}
\email{alec.hirschauer@morgan.edu}

\author[0000-0001-8658-2723]{Beena Meena}
\affiliation{National Radio Astronomy Observatory, 520 Edgemont Road, Charlottesville, VA 22903, USA}
\email{bmeena@nrao.edu}

\author[0000-0002-0577-1950]{Jeroen Jaspers}
\affiliation{Department of Physics, Maynooth University, Maynooth, Co. Kildare, Ireland}
\affiliation{Astronomy \& Astrophysics Section, School of Cosmic Physics, Dublin Institute for Advanced Studies, 31 Fitzwilliam Place, Dublin D02 XF86, Ireland}
\email{jeroen.jaspers@mu.ie}

\author[0000-0003-4023-8657]{Laura Lenki\'{c}}
\affiliation{IPAC, California Institute of Technology, 1200 East California Boulevard, Pasadena, CA 91125, USA}
\email{laura.lenkic@gmail.com}

\author[0000-0002-4663-6827] {R\'emy Indebetouw} 
\affiliation{University of Virginia Astronomy Department, P.O. Box 400325, Charlottesville, VA, 22904, USA}
\affiliation{National Radio Astronomy Observatory, 520 Edgemont Rd, Charlottesville, VA 22903, USA}
\email{remy@virginia.edu}

\author[0000-0001-5340-6774]{Karl D. Gordon} 
\affiliation{Space Telescope Science Institute, 3700 San Martin Drive, Baltimore, MD 21218, USA}
\email{}
\author[0000-0002-2449-0214]{Burcu G\"unay} 
\affiliation{Space Telescope Science Institute, 3700 San Martin Drive, Baltimore, MD 21218, USA}
\affiliation{Department of Physics \& Astronomy, Johns Hopkins University, 3400 N. Charles Street, Baltimore, MD 21218, USA}
\affiliation{Armagh Observatory and Planetarium, College Hill, Armagh, BT61 9DB, NI, UK}
\email{}
\author[0000-0003-3229-2899]{Suzanne Madden} 
\affiliation{Universit\'e Paris-Saclay, Universit\'e Paris Cit\'e, CEA, CNRS, AIM,91191 Gif-sur-Yvette, France}
\email{}
\author[0000-0002-5943-1222]{James Muzerolle Page} 
\affiliation{Space Telescope Science Institute, 3700 San Martin Drive, Baltimore, MD 21218, USA}
\email{}
\author[0000-0001-6576-6339]{Omnarayani Nayak} 
\affiliation{United States Naval Observatory, 3450 Massachusetts Avenue NW, Washington DC 20392, USA}
\email{}
\author[0009-0007-8087-6975]{Antonella Nota} 
\affiliation{Space Telescope Science Institute, 3700 San Martin Drive, Baltimore, MD 21218, USA}
\email{}
\author[0000-0003-4852-6485]{Theo Oneill} 
\affiliation{Center for Astrophysics |Harvard \& Smithsonian, 60 Garden Street, Cambridge, MA 02138, USA}
\email{}
\author[0000-0002-9573-3199]{Massimo Robberto} 
\affiliation{Space Telescope Science Institute, 3700 San Martin Drive, Baltimore, MD 21218, USA}
\affiliation{Department of Physics \& Astronomy, Johns Hopkins University, 3400 N. Charles Street, Baltimore, MD 21218, USA}
\email{}
\author[0000-0002-7759-0585]{Tony Wong} 
\affiliation{Department of Astronomy, University of Illinois at Urbana-Champaign, Urbana, IL 61801, USA}
\email{}
\author[0000-0002-9912-6046]{Petia Yanchulova Mercia-Jones} 
\affiliation{Space Telescope Science Institute, 3700 San Martin Drive, Baltimore, MD 21218, USA}
\email{}
%
%

\begin{abstract}

We present the first results from a JWST/NIRCam study of the star-forming region N159, one of the richest H{\sc ii} complexes in the Large Magellanic Cloud. This region stretches over $\sim$130 pc, and exemplifies 
large-scale sequential star formation. We present photometric results of $>$300,000 point sources combining data from ten near-infrared JWST filters spanning 1-4.7 microns. We detect sources as faint as $\sim$27 magnitudes at F115W, corresponding to stellar masses reaching below $\sim$0.1~M$_{\odot}$. By constructing several color-magnitude and color-color diagrams, we separate  young stars from main-sequence and older stars within N159, finding evidence for stellar populations younger than 0.5 Myr. Of the young population, we identify accreting, pre-main sequence stars from their Paschen-$\alpha$ (Pa$\alpha$) excess \textcolor{black}{and a population of sources bearing strong infrared (IR) excess, representing candidate young stellar objects (YSOs).}  Within this population, we identify a subpopulation of sources with H$_2$O ice absorption. The IR-excess-bearing YSO candidates correlate spatially with dust filaments and molecular gas demonstrating their embedded state. The distribution of stellar and \textcolor{black}{candidate} YSO populations across N159 reveal more evolved star-forming populations in the N159E region  and newer star formation in the N159W region, supporting the sequential star formation scenario for this region.

\end{abstract}

\keywords{\uat{Galaxies}{573} --- \uat{Large Magellanic Cloud}{903} --- \uat{JWST}{2291} --- \uat{Metallicity}{1031}}


\begin{figure*}[ht!]
\vspace{.25in}
\includegraphics[width=\textwidth]{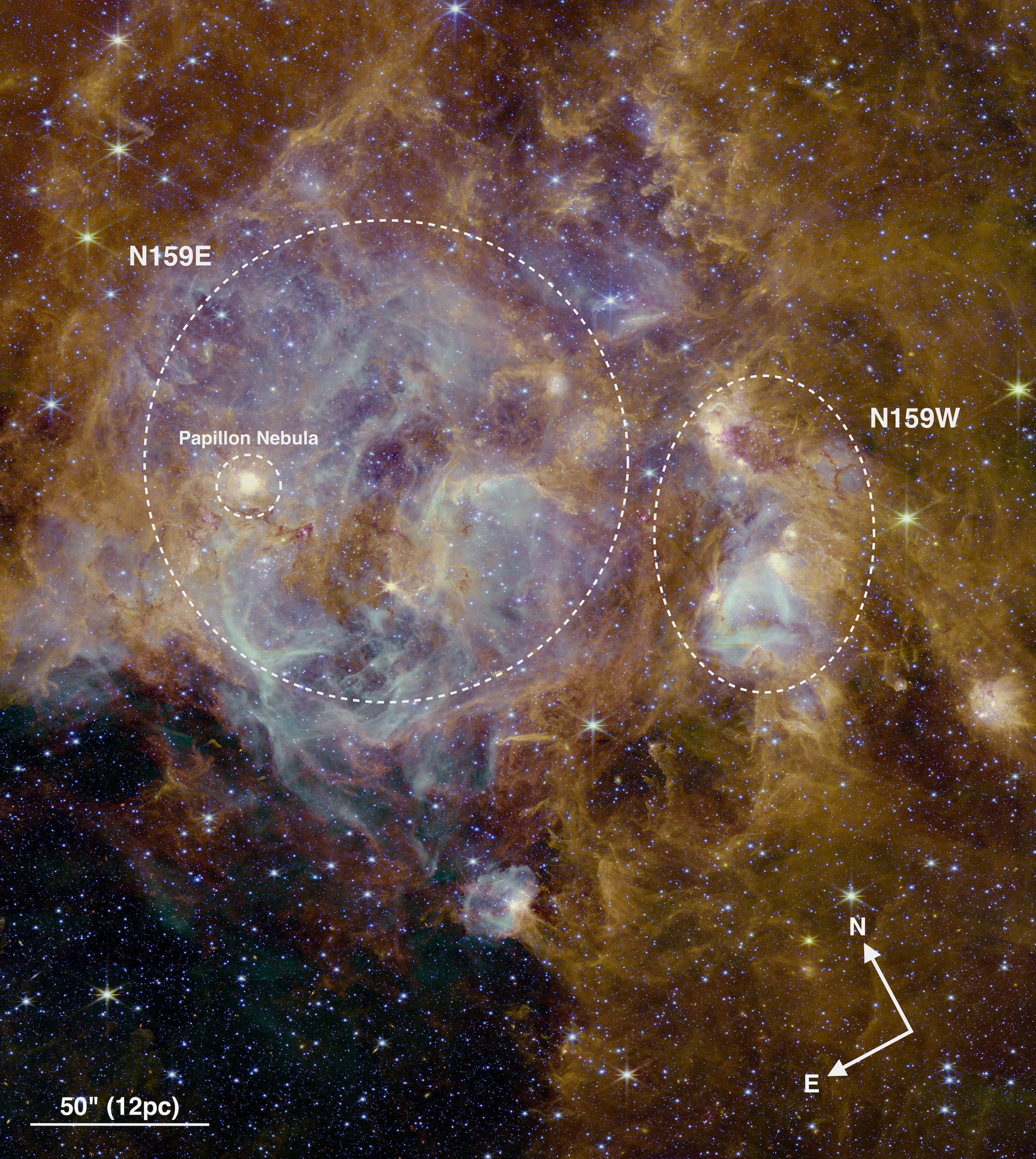}
\vspace{.1in}
\caption{The N159 star-forming complex as seen through six of JWST NIRCam's filters. Included filters and color key are as follows: 
\textbf{\textcolor{blue}{F115W}} (stellar photosphere and continuum emission),
\textbf{\textcolor{cyan}{F187N}} (Pa$\alpha$ emission),
\textbf{\textcolor{Green}{F300M}} (continuum emission and H$_2$O),
\textbf{\textcolor{YellowOrange}{F335M}} (PAH emission),
\textbf{\textcolor{red}{F460M}} (continuum emission and CO$_2$) and
\textbf{\textcolor{magenta}{F470N}} (H$_2$ emission).
}
\vspace{.1in}
\label{fig:six_color}
\end{figure*}

\section{Introduction}

The H{\sc ii} region N159 is one of the most recent ($\lesssim 3\, {\mathrm{Myr}}$, \citealt{bib:Jones_2005, bib:Kawamura_2009, bib:Minamidani_2011, bib:Saigo_2017}) sites of star formation (SF) within the Large Magellanic Cloud (LMC). N159 lies along a chain of giant molecular clouds (GMCs) spanning one kiloparsec and is situated just north of the quiescent ``molecular ridge", which contains a third of the total molecular mass of the LMC \citep{bib:Fukui_1999,bib:mizuno_2001}, and $\sim 600\, {\mathrm{pc}}$ south of 30 Doradus (the most luminous and most massive star forming region in the Local Group). Stretching over $130\, {\mathrm{pc}}$, N159 showcases sequential star formation on a large scale \citep{bib:Cohen_1988, bib:Fukui_1999, bib:Bolatto_2000,bib:Nakajima_2005, bib:Galametz_2013, bib:Bernard_2016, bib:Gordon_2017}.

The region is most evolved in its eastern side (N159E), which contains several OB stars \citep{bib:Meynadier_2004, bib:Saigo_2017} and hundreds of massive ($5-35\,  {\mathrm{M}}_{\odot}$) Herbig AeBe and young stellar objects (YSOs; \citealt{bib_Martin_Hernandez_2008, bib:Chen_2010, bib:Carlson_2012}). Strong emission from polycyclic aromatic hydrocarbon (PAH) molecules 
\citep{bib:Bolatto_2000, bib:Nayak_2018} is also present, along with water masers \citep{bib:Caswell_1981, bib:Oliveira_2006, 2010MNRAS.404..779E}.   A small ($\sim$ 2pc), high density, high excitation blob (HEB; \citealt{bib:Heydari_Malayeri_2010}), named the ``Papillon Nebula," is a precursor to ultra-compact \ion{H}{2} regions. The western part (N159W) is considerably less evolved, containing protostars more deeply embedded within molecular gas \citep{bib:Nayak_2018}. \cite{bib:Giannini_2013} detected [Fe {\sc ii}] emission in this region, suggesting the presence of protostellar jets or photoexcitation from external far ultraviolet (FUV) radiation (e.g. \citealt{bib:Mouri_2000}). This region has recently begun rigorous high-mass SF, akin to 30 Doradus, as evidenced by its high reservoir of 12CO(J = 3-2) (3 times that in 30 Doradus) and enhanced 12CO(J = 4-3/J = 1-0) ratio absent a well-defined H{\sc ii} region \citep{bib:Minamidani_2011}. The southern region (N159S) contains Herbig Ae/Be and reddened O-stars \citep{bib:Nakajima_2005} and Class O/I YSO candidates \citep{bib:Sewilo_2019} surrounding a pre-cluster core. It is the least evolved, with little ongoing star formation detected to date \citep{bib:Bolatto_2000, bib:Galametz_2013}. In this work, we will focus only on the N159E and N159W regions.

Metallicity in the LMC is notably low (0.5 Z$_{\odot}$; \citealt{bib:Westerlund_1997}), making N159 a unique (and accessible at 50 kpc; \citealt{bib:Feast_1999}) object of study to probe emerging SF in an environment significantly different from the Milky Way. Moreover, the impact of FUV radiation from O and B stars on the evolution of protoplanetary disks and the ISM can be investigated with the gradient in stellar ages and GMC morphologies in N159.

Past studies before the James Webb Space Telescope (JWST) have documented and investigated $\sim$200 very massive ($10- 40\, {\mathrm {M}}_{\odot}$) YSOs and embedded clusters, but lacked resolution and sensitivity in the visible and infrared (IR) wavelengths to identify lower mass objects \citep{bib:Chen_2010, bib:Carlson_2012, bib:Nayak_2018, bib:Reiter_2019}. JWST now allows for the detection of YSOs lower than $<1\, {\mathrm M}_{\odot}$ and the resolution of individual objects from compact clusters in nearby galaxies. Already, JWST has demonstrated these capabilities for star-forming regions in both the Small Magellanic Cloud (SMC) and LMC (e.g. 30 Doradus: \citealt{bib:Fahrion_2023, bib:Fahrion_2024}; NGC 346: \citealt{bib:Jones_2023}, \citealt{bib:Habel_2025}, \citealt{bib:Jaspers_2026}; N79: Jones et al. in prep.; NGC 602: \citealt{bib:Zeidler_2024}, \citealt{bib:Meena_2025}).

In this work, we present the first examination of the stellar and pre-stellar populations revealed within N159 by JWST. In \autoref{sec:obs_data} we describe the observations, data reduction and photometric extraction process. In \autoref{sec:pops} we describe our methods for identifying young and evolved populations. Finally, in \autoref{sec:spatial}, we discuss the spatial distribution of the various populations. We summarize our findings in \autoref{sec:conclusions}.

\begin{figure*}[ht!]
\centering
\includegraphics[width=.85
\textwidth]{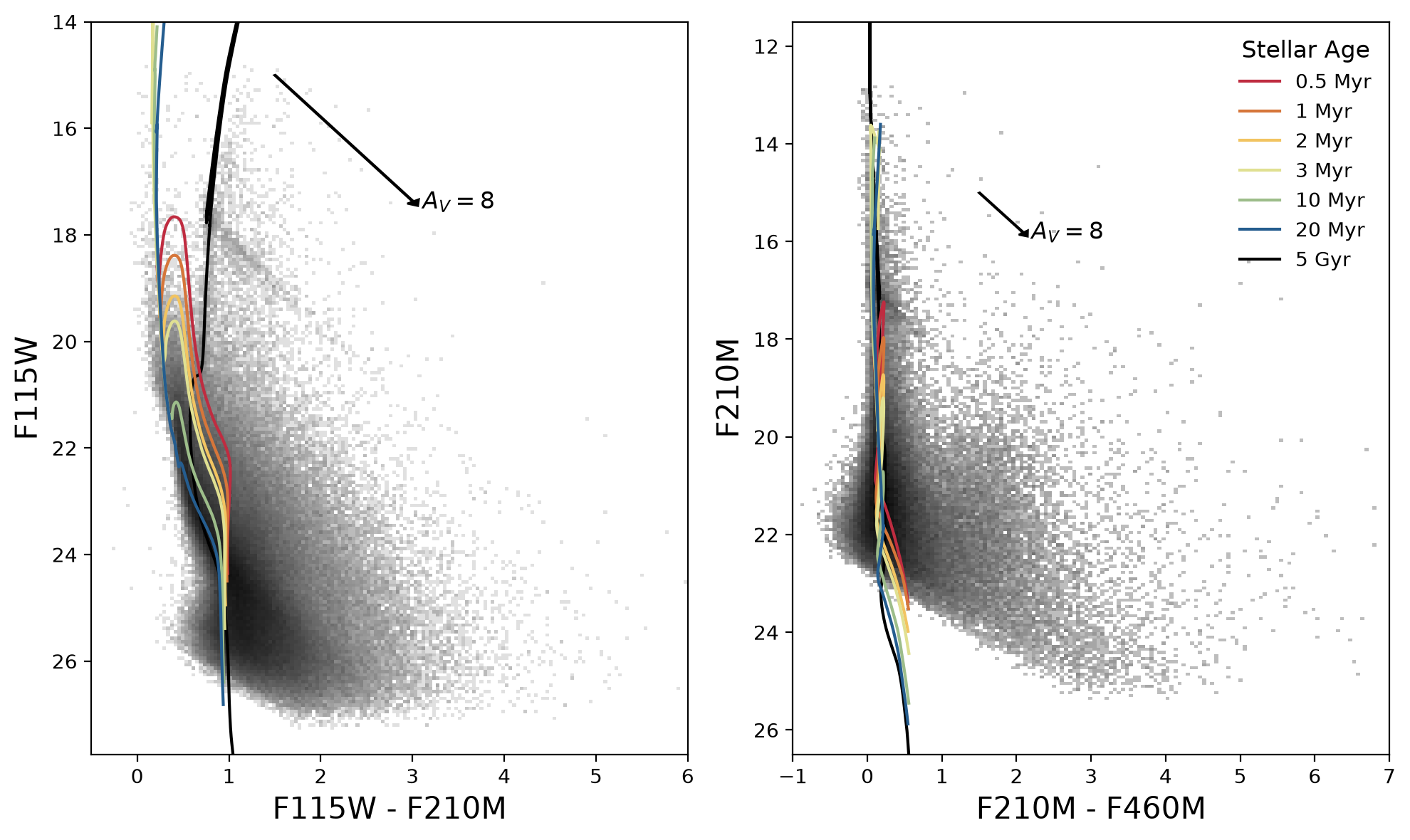}
\caption{CMDs of the sources detected within N159. Photometry for the entire field is shown in Hess format (gray). F115W-F210M (\textbf{left)} and F210M-F460M (\textbf{right)} CMDs with MIST isochrones from $0.5\, {\mathrm {Myr}}$ to $5\, {\mathrm{Gyr}}$ plotted across the entire photometric catalog. A reddening vector of A$_V$=8 is indicated by the arrow.}
\label{fig:CMD_ISO}
\end{figure*}

\section{Observations and Data Processing} 
\label{sec:obs_data}

\subsection{Observation Setup}\label{sec:obs_setup}
In this work, we examine near-IR imaging of the star-forming region N159 taken with JWST's NIRCam instrument \citep{bib:Rieke_2005,bib:Rieke_2023}. These observations were conducted as part of the JWST GO program $\#$5114 (PI: E. Sabbi). This paper focuses only on the set of observations within that program covering the main part of N159 (centered on R.A.=05:39:52.3800, Dec.=-69:45:42.50) which spans $\sim$35 arcmin$^2$ ($\sim 7300\, {\mathrm{pc}}^2$) and exposed for 3.74 hrs across ten filters: F115W, F140M, F187N, F210M, F212N, F300M, F335M, F360M, F460M and F470N. These data were taken as a 3x2 mosaic covering the chip and module gaps using the \textbf{4-SMALL-GRID-DITHER} sub-pixel dither pattern and the \textbf{Bright2} readout pattern with five groups, except for the F212N/F470N filter combination, which was increased to six. This observing setup was designed to attain \textit{S/N} $>$ 10 for all stars $\geq$ 0.2M$_{\odot}$ in all broad and medium filters.

\subsection{Image Processing}\label{sec:image_proc}

We processed the uncalibrated NIRCam images using version 1.18.0 of the JWST pipeline \citep{bib:Bushouse_2024}, Calibration Reference Data System (CRDS) version 12.1.5 and CRDS context \texttt{jwst\_1371.pmap}. After the standard Stage 1 and Stage 2 processing, we implemented an additional 1/\textit{f} noise correction step on the calibrated level 2 images using the \texttt{image1overf.py} routine from \cite{bib:1fcor}. To align our observations, we first created a template catalog based on the F300M imaging. Using the pipeline's Stage 3 \texttt{tweakreg} step, we aligned each exposure in this filter individually to the Gaia DRC catalog, using only sources with measured proper motion. We then invoked the Stage 3 step again, with \texttt{tweakreg} turned off, in order to assemble a complete, aligned mosaic with a preliminary source catalog. Finally, we repeated this two-step Stage 3 process with the remaining filters, matching the exposures to the now-Gaia-aligned F300M catalog, and creating background-corrected and skymatched mosaics.

\subsection{Color Images}\label{sec:color_image}
\autoref{fig:six_color} shows a false-color image of the N159 field, combining a selection of six of the ten filters (F115W, F187N, F300M, F335M, F460M and F470N). Colors have been assigned following wavelength, with bluer colors representing those at shorter wavelengths and redder colours representing longer wavelengths. Intricate nebulosity structure is apparent throughout the image, particularly in the northern and western regions. The southeastern region is relatively free of diffuse structure. Emission from the F335M filter, colored with gold, is persistent across the majority of the image, tracing dust structure via PAH emission at 3.3~$\mu$m. The bight ``Papillon Nebula" \citep{bib:Heydari-Malayeri_1999, bib:Jones_2005} H{\sc ii} region within N159E is resolved as a tight cluster of many distinct sources. Both the eastern and western regions are strongly illuminated by Paschen-$\alpha$ (\paa) emission (traced by the F187N filter in light blue). H$_2$ emission (seen as magenta in the F470N filter) closely traces the filamentary structure surrounding the east and west H{\sc ii} regions, and is particularly bright surrounding the northern region of N159W.

\subsection{Photometric Extraction and Correction}\label{sec:photom}
For extracting photometry, we used the Dolphot \citep{bib:Dolphin_2000,bib:Dolphin_2016} photometry package. \textcolor{black}{This widely-used, crowded field photometry package supports JWST's NIRCam module, performing PSF photometry incorporating JWST's WebbPSF models \citep{bib:weisz_2024}.} We performed source detection on the combined mosaics in order to detect the faintest possible sources. \textcolor{black}{To counter the nebulous background present within this region, we configured DOLPHOT to first mask out detected sources to create a master background of the field, mapping the nebulosity. This background was then subtracted from the original images, effectively performing local continuum subtraction. Finally, PSF fitting was performed on background-subtracted individual exposures.} 
We cleaned each single filter catalog of low-confidence sources (see \autoref{tab:cleaning}) before combining all filters via nearest-neighbor matching. Only sources detected in three or more filters were included in our final catalog. Finally, we converted our catalog from the AB to Vega magnitude system using the photometric offsets given by the CRDS reference file \texttt{jwst\_nircam\_abvegaoffset\_0002.asdf}. Our final catalog contains 352,703 point sources. Future works will make available and describe the photometry in further depth, including completeness limits.

\section{Population Analysis}\label{sec:pops}
In the following section, we discuss the stellar demographics revealed in N159. We examine the catalog described above using a combination of isochrones and color-magnitude diagrams (CMDs), identifying the color-space occupied by various populations.

\subsection{Stellar Ages and Isochrone Comparison}
\label{sec:isochrones}
We conduct CMD analysis using the ten JWST/NIRCam filters, making comparisons with isochrones to examine various populations within N159. We apply theoretical \textcolor{black}{MIST\footnote[1]{https://mist.science/}} isochrones \textcolor{black}{\citep{bib:dotter_2016, bib:choi_2016}} for the ages of 0.5, 1, 2, 3, 10 and 20~Myr to assess the young populations within the region, along with a 5~Gyr isochrone to identify the more evolved populations lying along the same line of sight. Because our catalog is not extinction corrected, we instead use isochrones with a foreground reddening of \textit{A$_V$ = 3} using \textit{R$_V$ = 3.1}\citep{bib:Meynadier_2004, bib:Jones_2005} which follows the extinction law of \cite{bib:Cardelli_1989}. We assume a metallicity value of [Fe/H]=-0.37 dex \citep{bib:Choudhury_2016}, and a distance modulus of 18.48 \citep{bib:Scowcroft_2016}.

\autoref{fig:CMD_ISO} shows the F115W-F210M \textcolor{black}{and F210M-F460M CMDs. Within the F115W-F210M CMD, the} F115W limit is $\sim$27 magnitudes. At approximately F115W=25, the corresponding mass along the 2~Myr isochrone is 0.1 M$_\odot$, demonstrating that we are sensitive to stellar and PMS sources within this mass range and below. We observe that the 5~Gyr isochrone aligns with the least reddened of the evolved red clump (RC) population, and closely matches the profile of the main sequence for magnitudes of $\sim$21 and greater. \textcolor{black}{In both CMDs shown in \autoref{fig:CMD_ISO}, a population of sources fainter and redder than the 0.5~Myr isochrone is apparent, containing significantly more IR-excess. In the F210M-F460M CMD in particular, we observe a population of reddened sources starting at $F210M-F460M\approx1.5$ and extending significantly past the elongated red clump, indicating the presence of an intrinsically young population distinct from evolved main-sequence sources which may be subject to foreground reddening by dust. We note, however, that the degree of extinction present in this region hinders precise age-dating of these young sources.}

\subsection{Main Sequence and Evolved Populations}
\label{sec:evolved_populations}
From the F115W-F210M vs F115W CMD, we can readily identify the color-space occupied by the main sequence and evolved stellar populations. The $0.5\, {\mathrm{Myr}}$ isochrone turnoff from the main sequence occurs at  F115W $\lesssim 18\,{\mathrm{mag}}$, beginning the upper main sequence (UMS). Within this population, we identify sources with F115W–F210M $<$ 0.3 as UMS stars. We flag those within the color-space F115W$<$19.5 and 0.6 $<$ F115W--F210M $<$ 1.5 as red giant branch (RGB) stars (see \autoref{tab:nircam_source_table}). The RC population is marked by an increase in source density within the RGB which begins at F115W$\approx$18 and extends along the direction of reddening. This population falls within a polygon as defined in \autoref{tab:nircam_source_table} spanning the magnitudes 17.2-20.0 that extends from a color of 0.5 to 2.0, beyond which the RC sources are comparable in density to those from general IR excess and difficult to differentiate. \textcolor{black}{We compare this extended RC to a reddening vector via the Python \texttt{dust\_extinction} package \citep{bib:Gordon_2024} and adopting the \texttt{G23} model \citep{bib:Gordon_2003, bib:Gordon_2023} with R$_V$=3.1. The discernible structure of the RC extends along the direction of the reddening vector to approximately A$_V$=8.}

\begin{figure*}[ht]
  \centering
  \begin{minipage}{0.4\textwidth}
    \includegraphics[width=\textwidth]{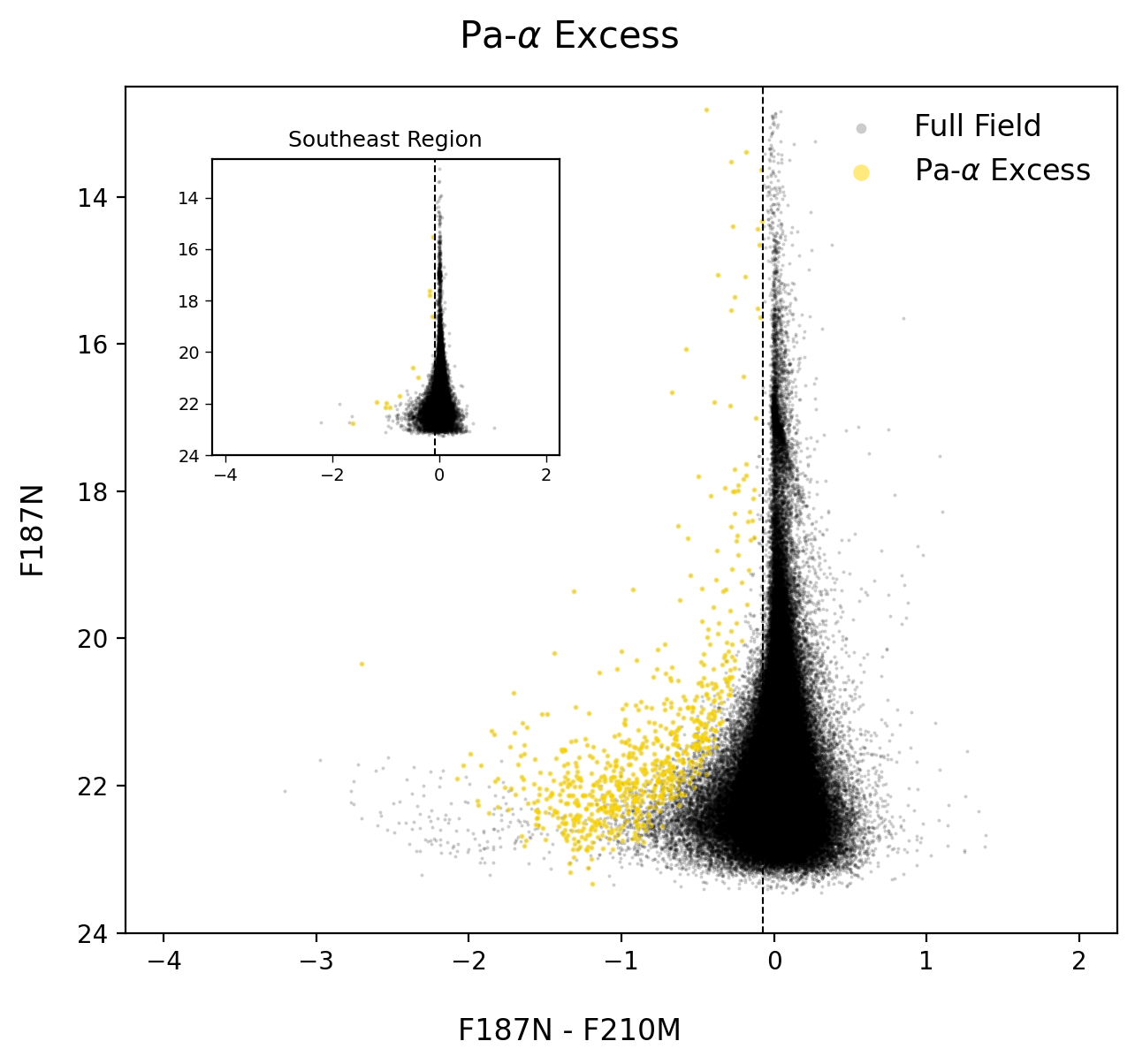}
  \end{minipage}
  \hspace{.3in}
  \begin{minipage}{0.4\textwidth}
    \includegraphics[width=\textwidth]{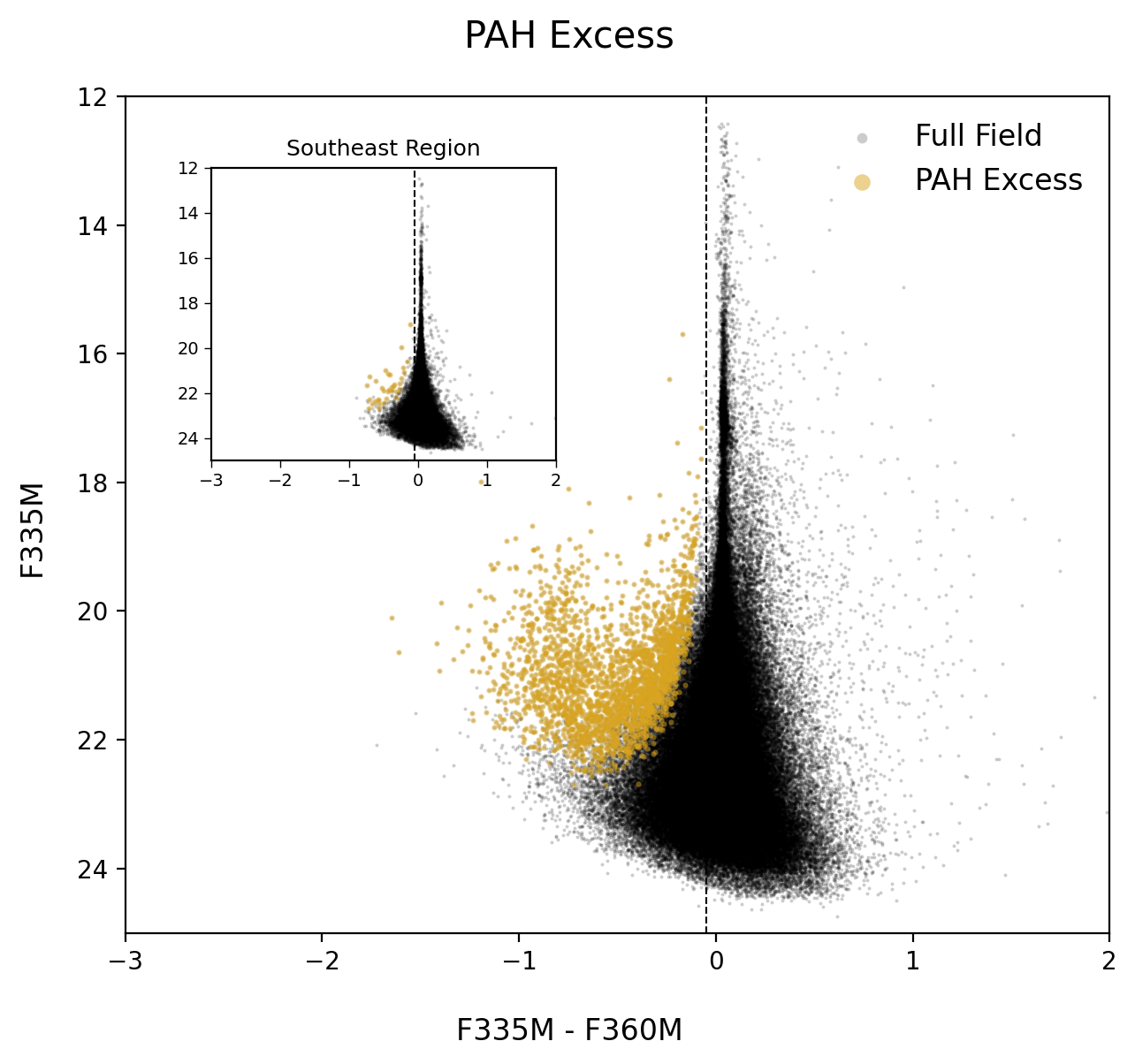}
  \end{minipage}

  \begin{minipage}{0.4\textwidth}
    \includegraphics[width=\textwidth]{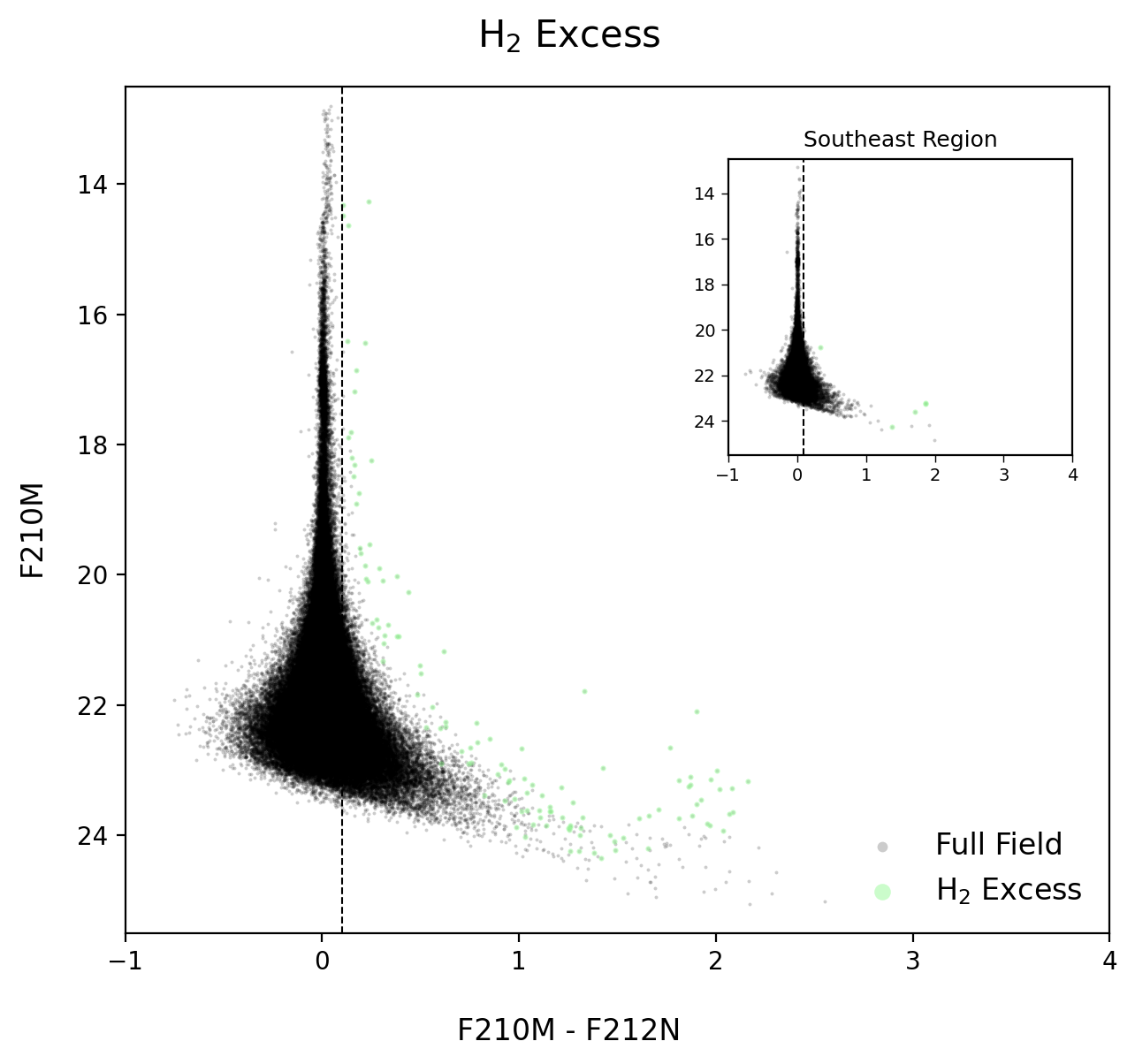}
  \end{minipage}
  \hspace{.3in}
  \begin{minipage}{0.4\textwidth}
    \includegraphics[width=\textwidth]{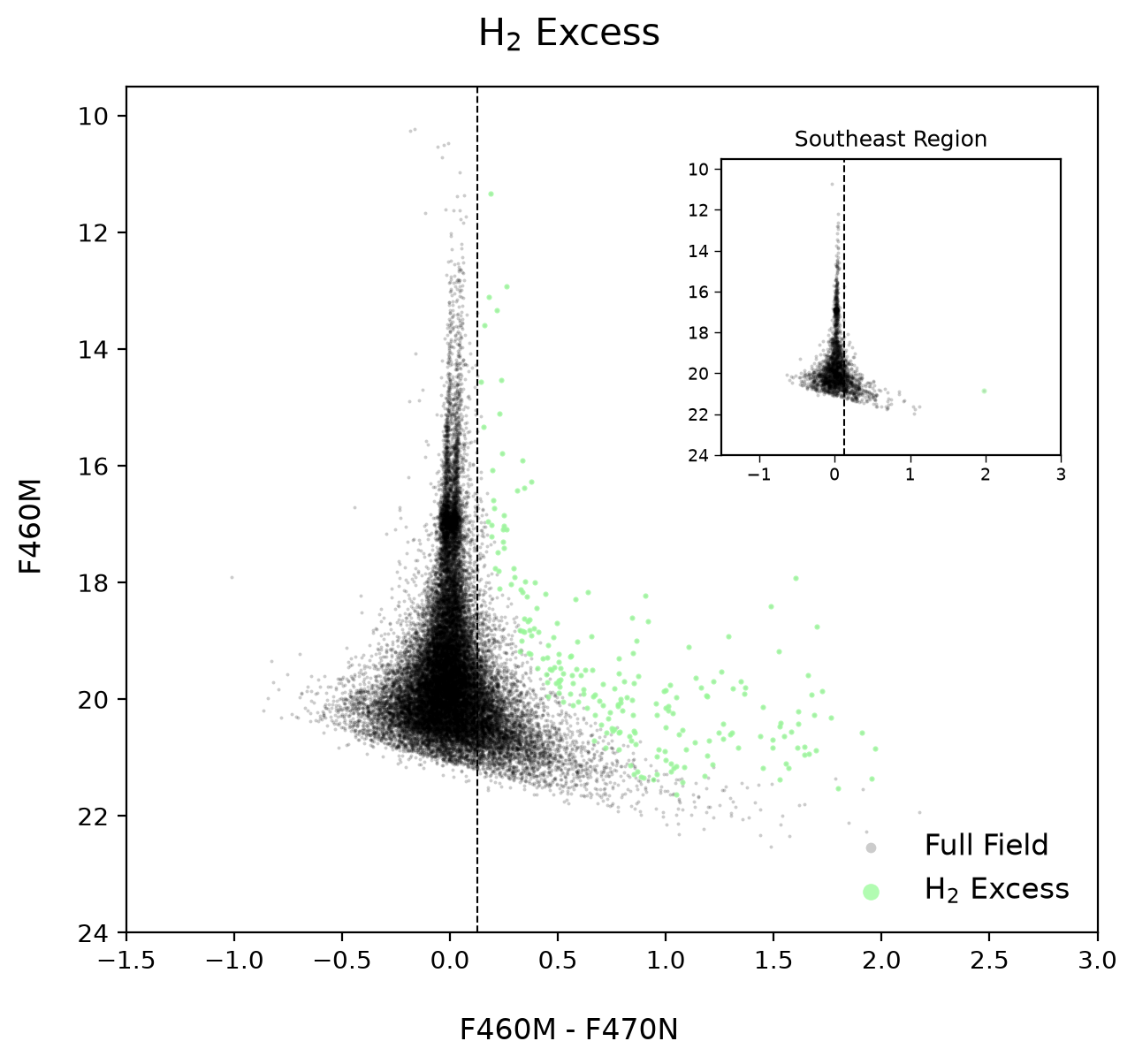}
  \end{minipage}

  \begin{minipage}{0.4\textwidth}
    \includegraphics[width=\textwidth]{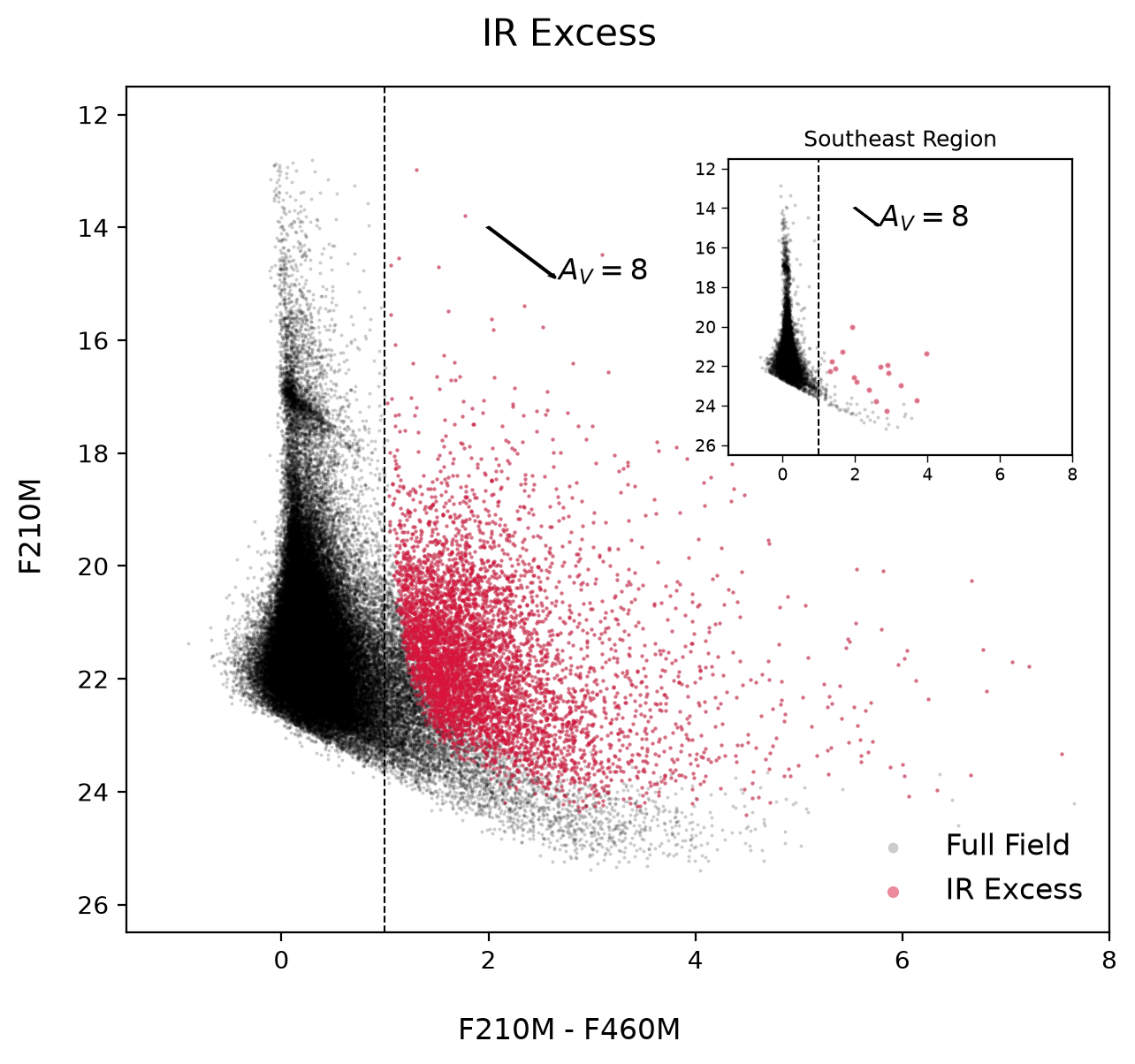}
  \end{minipage}
  \hspace{.3in}
  \begin{minipage}{0.4\textwidth}
    \includegraphics[width=\textwidth]{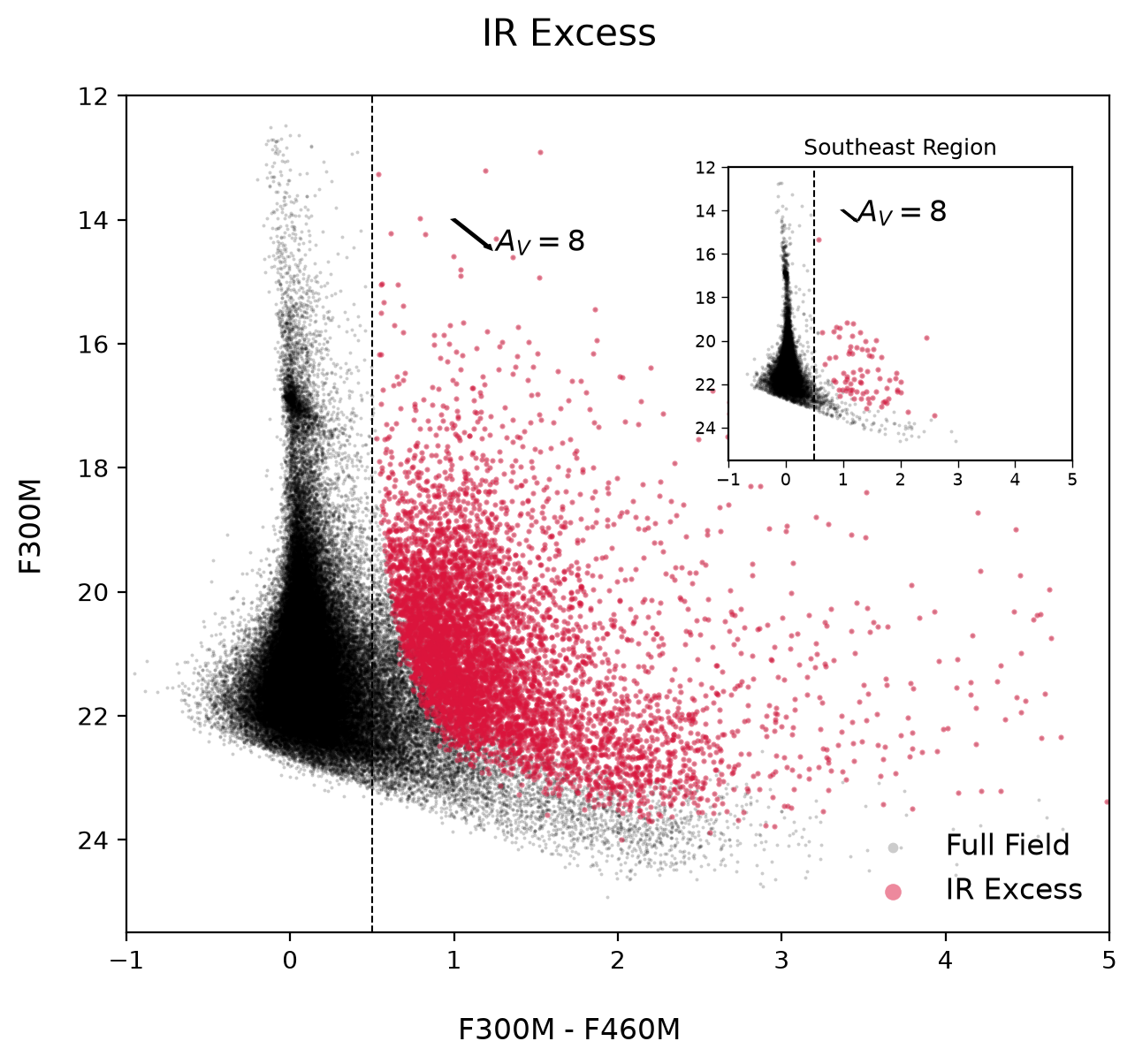}
  \end{minipage}

  \caption{CMDs showing color cuts defining several excess populations. \textbf{Inset in each plot}: the same CMD but for only the sources in the southeast region of the field containing mostly main sequence and evolved field stars. A reddening vector of A$_V$=8 is indicated by the arrow for the bottom row.}
\label{fig:inset_cmds}
\end{figure*}

\begin{figure*}[ht!]
\centering
\includegraphics[width=.8
\textwidth]{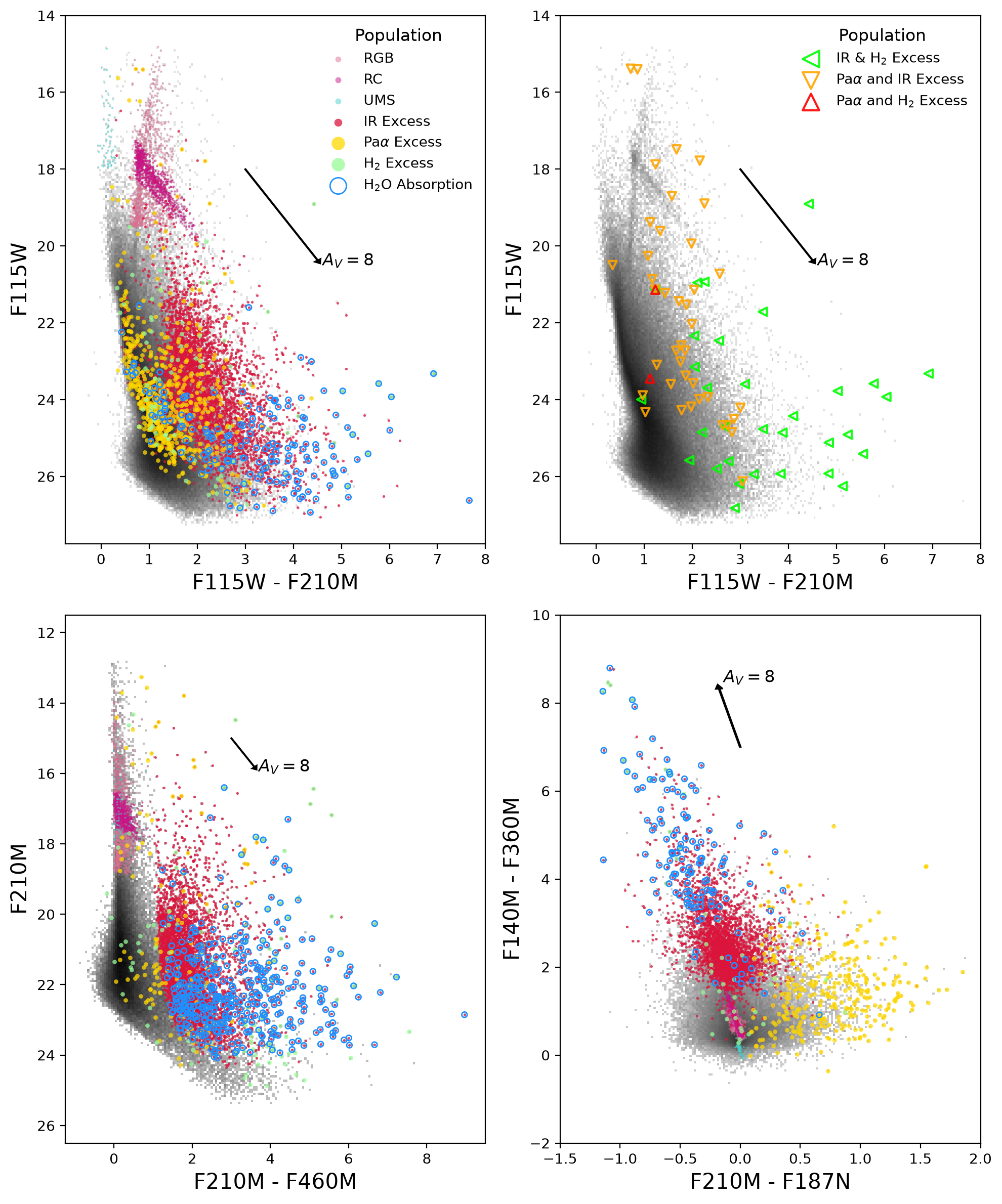}
\caption{CMDs and color-color diagrams (CCDs) of the stellar populations within N159. Photometry for the entire field is shown in Hess format (gray). 
\textbf{Top Left}: F115W-F210M CMD showing stellar populations and excess-bearing sources. 
\textbf{Top Right}: F115W-F210M CMD showing sources with multiple detected excesses. 
\textbf{Bottom Left}: F210M-F460M CMD showing the distribution of the redder, IR-excess bearing populations.
\textbf{Bottom Right}:F210M-F187N vs F140M-F360M CCD illustrating the distribution of \paa excess sources (x axis) and IR excess sources (y axis). 
A reddening vector of A$_V$=8 is indicated by the arrow in each subplot.}
\label{fig:CMD_POPS}
\end{figure*}

\begin{table*}[!htbp]
\centering
\begin{tabular}{|llr|}
\hline
Population & Color Criteria & Number of Sources   \\
\hline
UMS   				& F115W--F210M $< 0.3$ and F115W $< 18.0$                         & 71   \\ 
RGB  				& $0.6 <$ F115W--F210M $< 1.5$ and F115W $< 29.5$                 & 1866   \\ 
RC   				& inside (x1,y1 = (0.7, 18.2), x2,y2 = (0.7, 17.2),                & 1036    \\ 
                    & x3,y3 = (2.0, 19.0), x4,y4 = (2.0, 20.0)) where                &        \\
                    & x = F115W–F210M and y = F115W                                    & 	    \\
\hline
\paa ~Excess   		& F187N-F210M  $< -.075$                              	           & 726    \\ 
H$_2$ Excess 		& F210M--F212N $> 0.1$ or F460M--F470N $> 0.125$                   & 344    \\
PAH Excess    		& F335M--F360M $< -.05$ 		                                   & 2567   \\ 
IR Excess    	    & F210M--F460M $> 1.0$	  and F300M--F460M $> 0.5$                 & 5124   \\
H$_2$O Absorption   & Within IR Excess population and y $<$ 0.26x--0.19                & 389     \\
                    & where y = C$_{F210M,F330M,F360M}$ and  x = F210-F460M            &      \\
\hline
\hline
\bf{All Sources}  	& 							                                       & \bf{352,703}  \\ 
\hline
\end{tabular}
\caption{The stellar populations within N159 and selection criteria as identified using JWST/NIRCam. \paa, PAH, H$_2$ and IR excess sources must also pass the color test by 5 times the combined uncertainty from the two CMD filters.}
\label{tab:nircam_source_table}
\end{table*}

\subsection{Identifying Populations with Emission Excess via CMDs}
\label{sec:cmds-excess}
The narrow and medium band filters allow measurements of excesses by examining color-cuts within select CMDS. The F187N (\paa), F212N (H$_2$) and F470N (H$_2$) filters can be used in conjunction with filters sensitive to the continuum to identify the presence of specific narrow line emission. Similarly, the F335M filter is sensitive to fluorescing PAH emission in heated dust via C–H bond stretching \citep{bib:Rodriguez_2023, bib:Dale_2023, bib:Sandstrom_2023}.

In general, we were able to identify the color-space in which the \paa, PAH, H$_2$, and IR excesses in relevant CMD combinations occur by comparing CMDs created from subsets of sources within different regions of our field. The southeast quadrant of our field of view is relatively free of diffuse nebulosity (see \autoref{fig:six_color}). As seen in \textcolor{black}{the inset plots within} \autoref{fig:inset_cmds}, CMDs created from sources in this region are populated mainly by a clearly-defined main-sequence and an evolved population, and show little emission line or IR excess. With these southeast region CMDs as reference, we were able to determine the color-space defining excess-bearing populations.

\autoref{tab:nircam_source_table} contains the list of our color cuts illustrated in \autoref{fig:inset_cmds}. In order to account for photometric uncertainties, we required that each source have a magnitude error of less than 0.1, except for the F187N, F212N and F470N bands, where we relaxed this requirement to 0.25. Following a similar approach to \cite{bib:Fahrion_2024}, we considered only those sources that surpassed the color cuts with a statistical significance of 5$\sigma$ or greater, where $\sigma$ is defined as their combined uncertainty between the two CMD filters. This approach allows us to identify sources displaying excess, while accounting for growing photometric errors as magnitude increases.

We considered the F187N--F210M CMD to identify sources with excess from the \paa~emission line at 1.878~\micron. Note that in this CMD, excess emission from hydrogen recombination results in colors less than zero. Both the F212N and F470N filters are sensitive to excess due to molecular hydrogen emission (H$_2$) from the H$_2$~1-0~S(1) and H$_2$~0-0~S(9) lines, respectively. We used F210M--F212N and F460M--F470N CMDs to identify sources showing this excess, flagging sources that pass the color cut in one or both CMDs. PAH excess was identified by the F335M-F360M CMD. For flagging sources with IR excess, indicative of a rising near-IR spectral energy distribution (SED), we considered a combination of two CMDs, F210M--F460M and F360M--F460M. Inclusion of the F210M filter allows us to consider the same color-space potentially occupied by sources with H$_2$ or \paa~excess, however we note that demanding a detection in the F210M filter may eliminate sources that are bright toward mid-IR wavelengths, but undetected at 2.1~$\mu$m. 

\textcolor{black}{Additionally, as we are interested in identifying sources bearing excess IR emission that may be due to thermal emission from an envelope or disk rather than from reddening via dust along the line of sight, we define our color-selections to flag only sources with a color redder than where the maximum reddening measured from the RC (A$_V$ $\cong$ 8) would move a source from the main sequence. In this way, we reduce the degeneracy between highly-dust-reddened main sequence or evolved sources and young sources bearing circumstellar disks and envelopes at the expense of including more-evolved YSO sources which bear lesser IR-excess. As the extinction law flattens toward the mid-IR, (e.g. 4.5–8.0 $\mu$m, \citealt{bib:lutz_1999, bib:indebetouw_2005, bib:Hartmann_2005, bib:Flaherty_2007, bib:Gutermuth_2008}), continuing studies of this region will incorporate photometry obtained from JWST's Mid-Infrared Instrument (MIRI) imager, which is less impacted by differential reddening, to further disentangle this degeneracy and to differentiate YSOs by evolutionary Class.}

\textcolor{black}{In total we find 5124 sources satisfying our IR-excess color criteria. A comparison of these sources with those showing narrow-line excess reveals 48 sources with both IR and \paa excess. In the top right plot in \autoref{fig:CMD_POPS}, these sources tend to be bright within the F115W-F210M CMD and are redder than the main sequence, possibly representing massive later-stage YSOs. Conversely, those sources with only a \paa excess detection follow the main sequence more closely to the CMD (see top left CMD in \autoref{fig:CMD_POPS}), and are fainter on average, likely comprising a population of more-evolved YSOs or PMS stars. We observe a 58 sources with IR and H$_2$ excess. Three sources exhibit both \paa and H$_2$ excess, with one also showing IR excess.}

\subsection{H$_2$O Absorption in IR Sources}
\label{sec:h2o}

The circumstellar environments of YSOs are known to be home to rich organic chemistry (e.g. H$_2$O, CO and CO$_2$ ices). These, along with more complex molecules (CH$_3$OH, CH$_3$OCHO; \citealt{bib:Pontoppidan_2008, bib:Nayak_2024}) are key building blocks for life \citep{bib:Ehrenfreund_2000, bib:Fedoseev2017}.

Water ice is often seen in absorption at 3.05~$\mu$m in young YSOs. We use the F300M filter, centered on this feature, to search for evidence of H$_2$O absorption in sources identified with IR excess, that is, those that are also candidate YSOs. To do this, we employ a pseudo-two-color diagram (or ``chromosome map"), adapting the methodology from \cite{bib:Milone_2017a, bib:Milone_2015b}. Because the continuum slope of a YSO's SED may be increasing or decreasing across 3.05~$\mu$m depending on its evolutionary state (e.g. \citealt{bib:Furlan_2016}), the F210M-F300M and F300M-F360M CMDs cannot reliably distinguish H$_2$O ice in absorption based on a color excess or deficit. We instead construct the pseudo-color $C_{F210M,F300M,F330M}=(F210M-F300M)-(F300M-F360M)$. Functionally, this pseudo-color approximates the difference between the continuum spanning the 3.05$\mu$m feature and the magnitude measured by the F300M filter. For example, for a YSO with a flat continuum from $\sim$2.1$\mu$m--3.6$\mu$m (e.g. F210M~=~F360M), $C_{F210M,F300M,F330M} < 0$ would indicate absorption at 3.05~$\mu$m.

\begin{figure}[ht!]
\centering
\includegraphics[width=\linewidth]{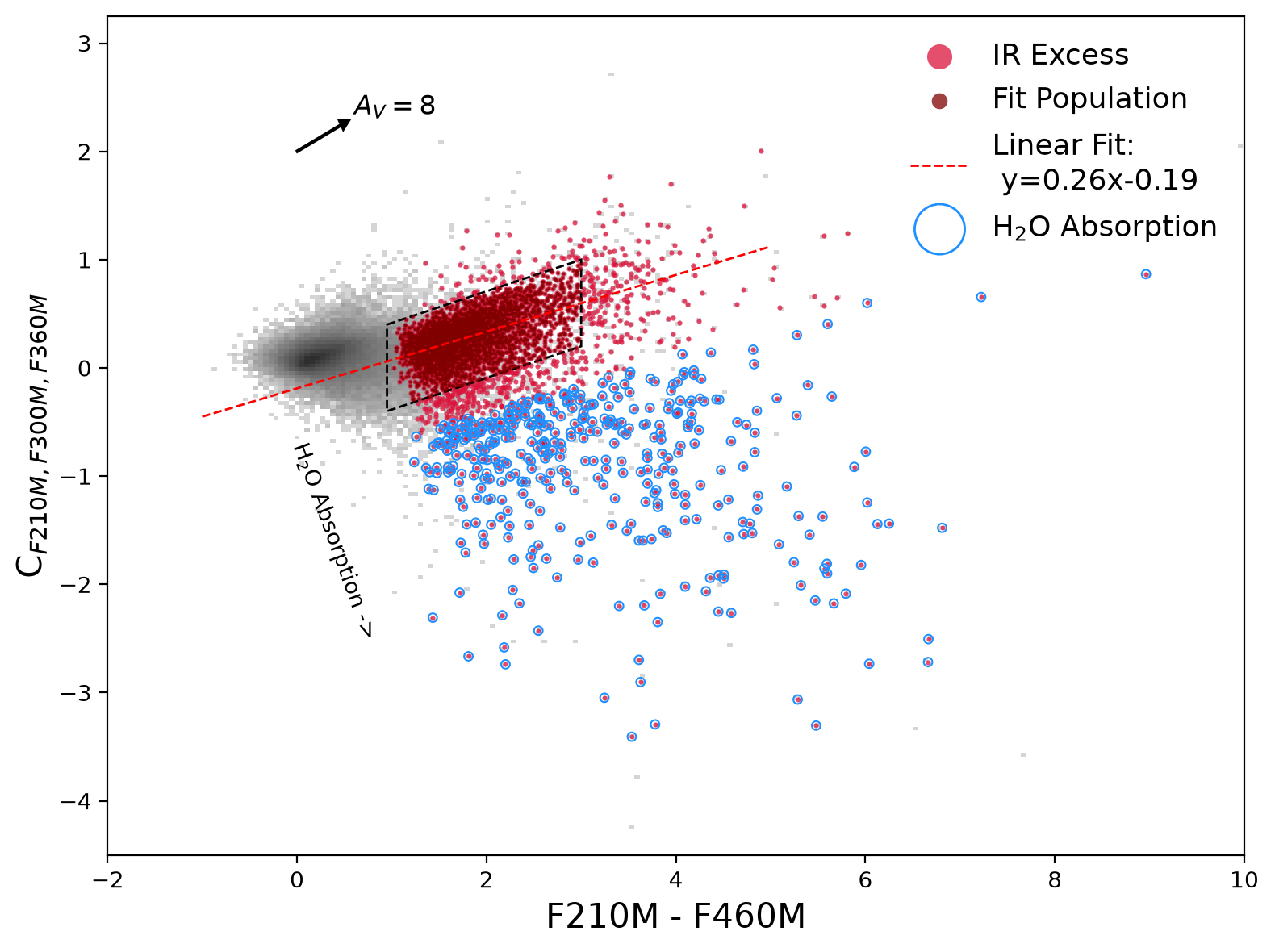}
\caption{F210M-F460M vs C$_{F210M,F300M,F360M}$ “chromosome map” (see text) showing sources within the IR excess population with evidence for H$_2$O absorption in the F300M band. The dashed red line shows the linear fit to the IR population within the black, dashed box. Sources flagged with H2O absorption fall 0.75 magnitudes below the red dashed line.}
\label{fig:H2O_Chrome}
\end{figure}

In \autoref{fig:H2O_Chrome}, we create a chromosome map to identify IR excess sources with strong evidence of water ice absorption by plotting the $C_{F210M,F300M,F330M}$ pseudo-color against F210M-F460M, a longer-baseline color which is insensitive to water ice absorption. We observe that the bulk of the IR excess sources form an extended population sloping upwards, as well as a scattered population extending to $C_{F210M,F300M,F330M}\approx-3.5$. Assuming that the bulk of the IR excess sources have little ice absorption, we find a linear fit to a subset of the IR sources composing the region of least scatter, defined by the polygon 
\textcolor{black}{x1,y1 = (0.95, 0.4), x2,y2 = (3.0, 1.0), x3,y3 = (3.0, 0.2), x4,y4 = (0.95, -0.4),}
where x = F210M-F460M and y = $C_{F210M,F300M,F330M}$. This region encompasses \textcolor{black}{3760 of the 5124} IR excess sources. Finally, we define an arbitrary cutoff of 0.75 magnitudes below this line, and flag IR excess sources in this region as candidates with detected water ice in absorption. In total, we identify \textcolor{black}{389} candidate objects. In theory, a greater proportion of the IR excess sources may also contain water absorption; however, in this work we focus on those with the strongest evidence. 

The sources we identify extend to our lower mass detection limit  
and thus likely represent the lowest-mass candidate YSOs with near-IR ice detected in the LMC to date. NIRSpec spectroscopy of these sources would be needed to precisely measure the column density of H$_2$O ice. Recent mid-IR spectroscopic studies of SMC and LMC YSOs with JWST have revealed ices with complex organics \citep{bib:Nayak_2024, bib:Habel_2025}. With JWST NIRSpec, \cite{bib:demarchi_2025}  measured H$_2$O ice in low-mass ($\sim$2~M$_\odot$) SMC YSOs. Importantly, this work also found that the relative ice abundances (e.g. column density ratios of H$_2$O, CO$_2$ and CO) did not significantly differ from those in massive Milky Way YSOs, suggesting metallicity may not significantly impact ice chemistry.  Followup spectroscopy of our N159 YSOs \textcolor{black}{candidates} with photometrically detected H$_2$O ice absorption would provide critical insights into the relative ice abundances at lower stellar masses than have been previously studied in the LMC.

\begin{figure*}[ht]
  \centering
  \includegraphics[width=\textwidth]{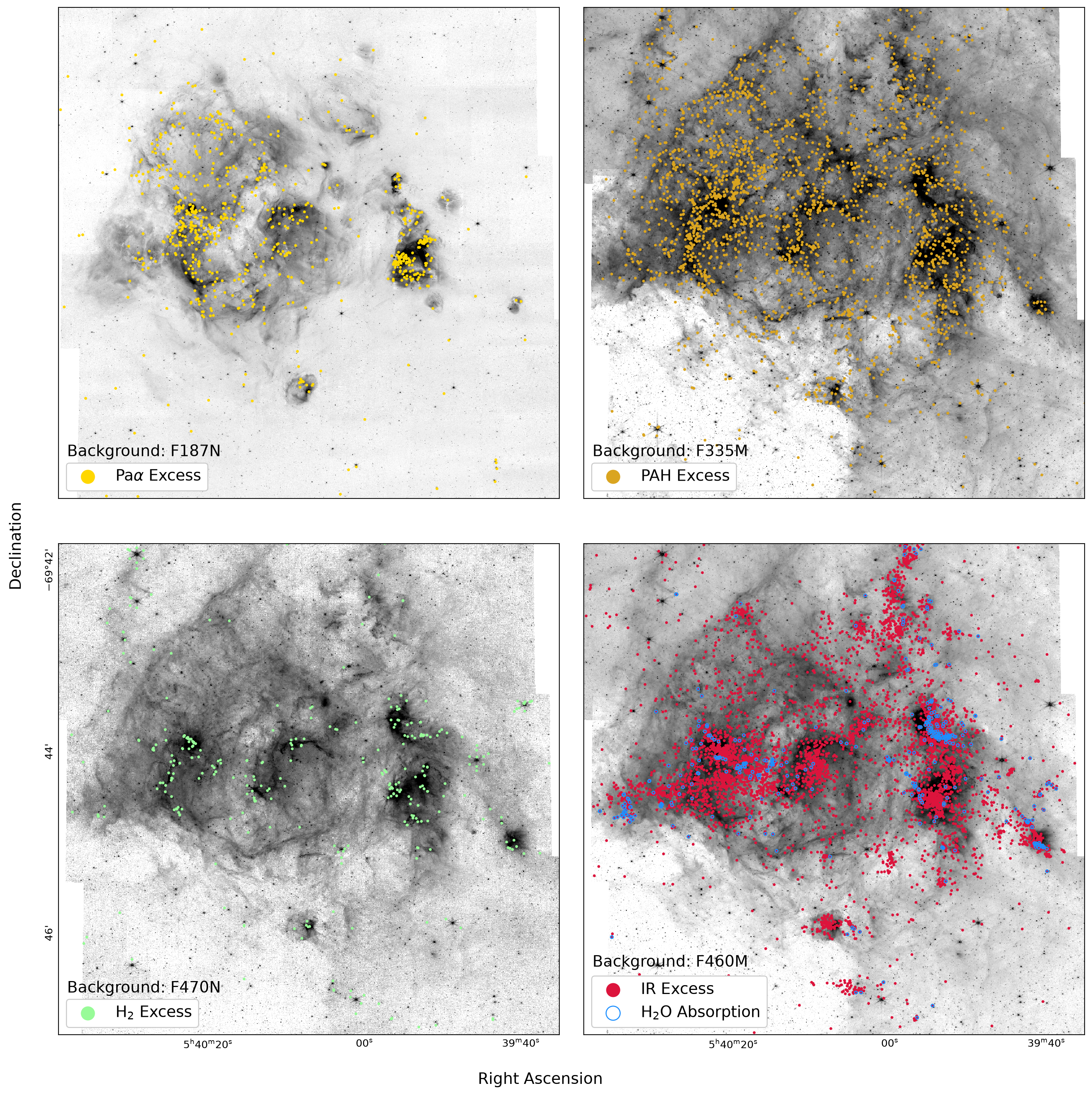}
  \caption{The spatial distributions of populations with identified excess emission or H$_2$O ice absorption. }
\label{fig:distribution4}
\end{figure*}

\begin{figure*}[ht]
  \centering
  \includegraphics[width=\textwidth]{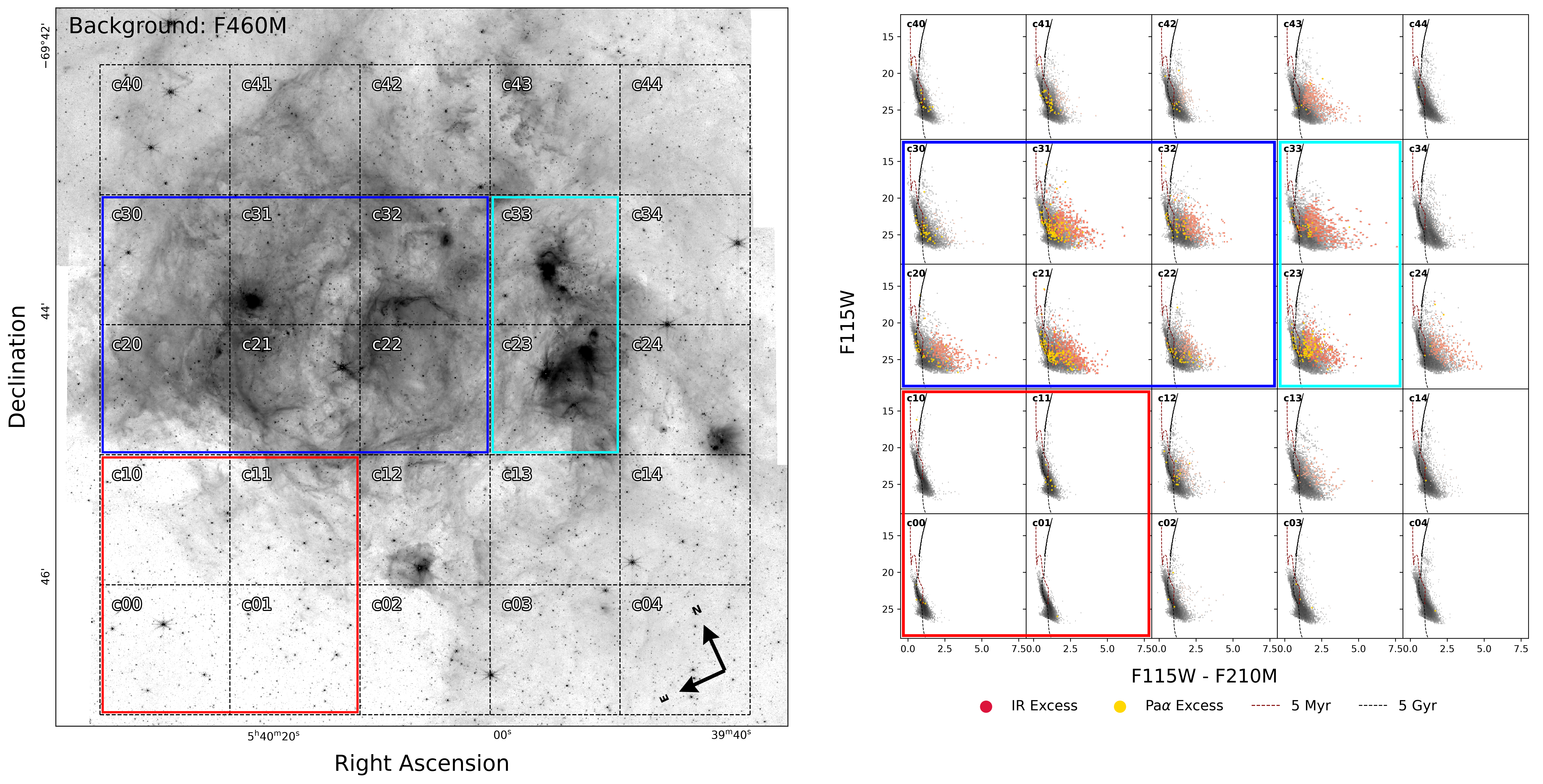}
  \caption{\textbf{Left}: NIRCam F460M imaging of N159 subdivided into 25 regions. Cells encompassing N159E and N159W are outlined in blue and cyan, respectively. The southeast cells used to guide CMD color selections are outlined in red. \textbf{Right}: F115W-F210M vs F115W CMDs showing the stellar populations in the corresponding 25 regions. The maroon and black isochrones trace 5~Myr and 5~Gyr, respectively. Sources with IR excess and \paa excess are marked in red and yellow, respectively.}
\label{fig:distribution2}
\end{figure*}

\begin{figure*}[ht!]
\includegraphics[width=.9\textwidth]{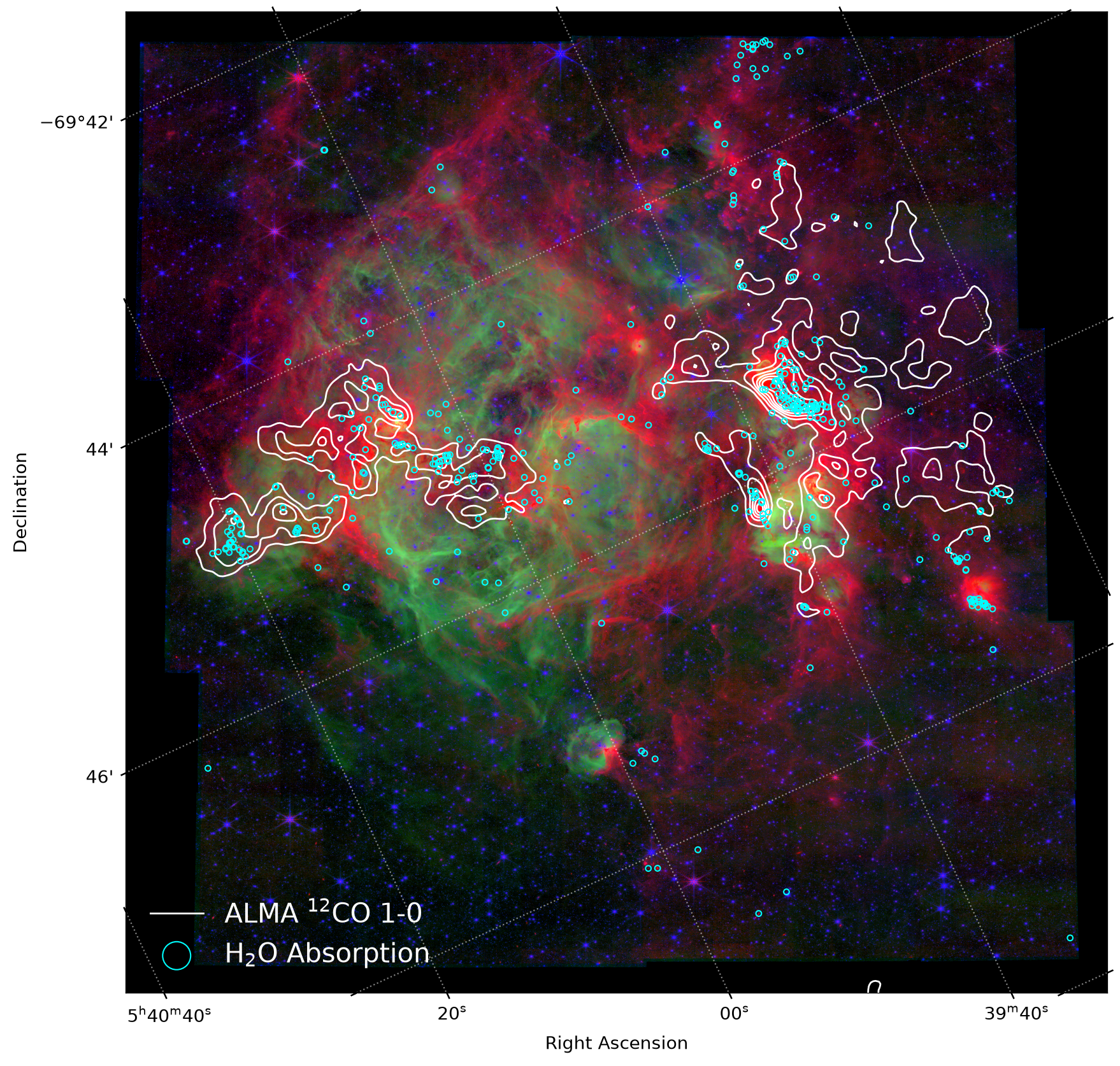}
\vspace{-.1in}
\caption{A three-color image of the N159 region with ALMA 12CO(J=1-0) contours over plotted and sources with H$_2$O absorption marked.
Included filters and color key are as follows: \textbf{\textcolor{blue}{F115W}} (stellar photosphere and continuum emission),
\textbf{\textcolor{Green}{F187N}} (Pa$\alpha$ emission), \textbf{\textcolor{red}{F470N}} (H$_2$ emission).
}
\label{fig:h2o_alma}
\end{figure*}

\section{Spatial Distribution of Populations}\label{sec:spatial}

In this section, we discuss the spatial distribution of the different populations across N159 and with respect to the dusty and nebular emission observed in our NIRCam images. \autoref{fig:distribution4} shows the spatial distribution of the populations with identified \paa, H$_2$, PAH or IR excess as described in \autoref{sec:cmds-excess} against the F335M mosaic which closely maps the dust structure in the region. 

Plotting sources with \paa excess (that is, likely PMS stars or YSOs with disks) immediately reveals that they trace the eastern and western star-forming regions closely. We also observe that sources in the western region are clumped more tightly than in the east, where they are more dispersed. \textcolor{black}{The left panel of \autoref{fig:distribution2} shows N159 split into subregions, and the right panel shows the F115-F210M CMD for each.} A manual inspection of where these sources lie in the F115W-F210M CMDs reveals that both regions contain at least some \textcolor{black}{\paa excess} sources close to the main sequence, indicating that dust is not excessively reddening this population. However, sources in the western region \textcolor{black}{(e.g in subregion c23)} show preferentially more IR excess, whereas more sources in the eastern \textcolor{black}{(e.g in subregions c31 and C21)} region lie closer to, or on top of the main sequence. This, along with the more dispersed distribution of the eastern \paa excess sources supports the notion of sequential star formation occurring across N159, with the most recent epoch occurring in the western region where the \paa excess sources represent PMS stars and younger YSOs.

H$_2$ excess sources (potentially tracing jets and disk fluorescence from extreme ultraviolet radiation) are concentrated toward the centers of N159E and N159W, but are less dispersed overall than the \paa excess sources. Notably, the cluster of stars composing the ``Papillon Nebula" in N159E is a hotspot for H$_2$ excess sources. 

Sources with PAH excess are closely associated with the dusty filaments surrounding the H{\sc ii} regions. In the F335M mosaic image, all but the southeast region show general elevated emission from diffuse nebulosity. We note that the general diffuse emission is not itself associated with an increased abundance of PAH sources, confirming that our PSF photometry is successfully detecting PAH emission in the immediate environment of each source.

The IR excess sources (containing candidate YSOs) follow a similar spatial distribution to the PAH excess sources, being distributed generally along the dusty filaments. This follows the understanding that star-formation occurs in highly-embedded regions rich in dust and molecular gas. However, unlike the PAH sources, we also observe an abundance of IR sources in the center of the H{\sc ii} regions in N159E and N159W.  

\textcolor{black}{As seen in \autoref{fig:distribution2}}, the bottom left region (c00, c01, c10 and c11) shows the CMDs of the non-star-forming background population, which have little to no IR excess sources. Both the western region (c23, and c33) and the eastern region (c20, c21, c22, c30, c31 and c32) reveal prominent IR excess sources in their CMDs, indicating active star formation. However, the western region shows a greater density of sources with a large amount of IR excess in comparison to the eastern region.  This suggests that the generally-redder western sources represent younger, more embedded YSO candidates than the eastern sources, supporting the scenario of sequential star formation in the region.

In \autoref{fig:distribution4}, the H$_2$O absorption sources are overlaid on top of the IR excess sources. These objects potentially represent the most embedded and youngest YSOs. These H$_2$O absorption sources are clustered tightly together in comparison to the IR excess sources and align more closely with dust filaments.  In the eastern region, the H$_2$O absorption sources  are more clustered compared to the \paa sources, tracing the locations of past star formation. The densest cluster appears in the western region along an arced filament. \textcolor{black}{In figure \autoref{fig:h2o_alma} we show ALMA $^{12}$CO(J=1-0) contours (ALMA projects 2012.1.00554.S and 2016.1.00782.S) overplotted on a three color image of our region with the H$_2$O-bearing sources marked. We see that the arc-like structure formed of a chain of sources is coincidental with an area of high CO column density. This ridge was identified as one of the most massive protocluster systems in the Local Group, by \cite{bib:Tokuda_2022} who theorized that it represents the site of a  large-scale gas compression event which triggered abundant star formation. In general, we observe that H$_2$O-bearing bearing sources follow closely the distribution of the CO contours, indicating an enriched reservoir of molecular compounds accessible during the formation of the YSOs within these regions, or an enhanced density of H$_2$O molecules co-located with the CO gas in the interstellar environment along the line of sight toward these sources.}

\section{Conclusions}\label{sec:conclusions}

\begin{enumerate}
\item Using JWST's NIRCam instrument, we measure PSF photometry for more than 300,000 sources present in at least three NIRCam bands in the LMC star formation region N159. We detect sources as faint as $\sim$27 magnitudes in the F115W band, extending to stellar masses below $\sim$0.1 M$_{\odot}$.

\item We compared CMDs created from the F115W, F210M and F460M filters to a range of theoretical isochrones. 
We find close agreement with the photometry and identify a significant population younger than 0.5~Myr.

\item Using CMDs involving NIRCam's ten filters, we identify four subpopulations of PMS and YSO candidates (IR excess emission sources),  by their excess emission in  \paa ,  H$_2$, and PAH and their absorption in H$_2$O due to water ice.  These subpopulations have different spatial distributions, indicating a sequential star formation process in the region.

\item The N159W region has a greater population of younger YSO candidates than the N159E region, more compact clusters of H$_2$O-absorption and H$_2$-excess sources and fewer \paa-excess sources. Our observations corroborate prior studies that N159W region is a younger, less evolved star formation region than N159E. 

\end{enumerate}

\begin{acknowledgments}
This work is based on observations made with the NASA/ESA/CSA James Webb Space Telescope. The data were obtained from the Mikulski Archive for Space Telescopes at the Space Telescope Science Institute, which is operated by the Association of Universities for Research in Astronomy, Inc., under NASA contract NAS 5-03127 for {\em JWST}. These observations are associated with program \#5114.
Some/all of the data presented in this paper were obtained from the Mikulski Archive for Space Telescopes (MAST) at the Space Telescope Science Institute. The specific observations analyzed can be accessed \textcolor{black}{via~\dataset[10.17909/0zvs-k944]{http://dx.doi.org/10.17909/0zvs-k944}.}

This work makes use of the following ALMA data: ALMA$\#$ 2012.1.00554.S, 2016.1.00782. 

NH and MM acknowledge that a portion of their research was carried out at the Jet Propulsion Laboratory, California Institute of Technology, under a contract with the National Aeronautics and Space Administration (80NM0018D0004).  NH and MM acknowledge support through STScI/{\em JWST} grant 5114.016.
JJ acknowledges support from the Research Ireland Pathway programme under Grant Number 21/PATH-S/9360.
O.C.J. acknowledges support from an STFC Webb fellowship.
\end{acknowledgments}




%
\facilities{JWST(NIRCam), ALMA}

\software{astropy \citep{2013A&A...558A..33A,2018AJ....156..123A,2022ApJ...935..167A},
            DOLPHOT \citep{bib:Dolphin_2000,bib:Dolphin_2016},
            \texttt{image1overf.py} \citep{bib:1fcor}}


\appendix
\label{sec:appendix}
In this appendix we present a table listing the parameters used to clean low-confidence sources from our photometric catalog.

\setcounter{table}{0}
\renewcommand{\thetable}{A\arabic{table}}

\begin{table}[h]
    \centering
    \begin{tabular}{|l|cccccccccc|}
    \hline
         & \textbf{F115W} & \textbf{F140M} & \textbf{F187N} & \textbf{F210M} & \textbf{F212N} & \textbf{F300M} & \textbf{F335M} & \textbf{F360M} & \textbf{F460M} & \textbf{F470N} \\
\hline    
    \texttt{e\_mag\_lim    } &  10.0  & 10.0&  10.0&  10.0&  10.0&  10.0&  10.0&  10.0&  10.0& 10.0\\
    \texttt{mag\_lim    } &  40.0  & 40.0&  40.0&  40.0&  40.0&  40.0&  40.0&  40.0&  40.0& 40.0\\
    \texttt{round\_lim    } &  1.0  & 2.0&  2.0&  2.0&  2.0&  2.0&  2.0&  3.0&  3.0& 3.0\\
    \texttt{sharp\_lo\_lim} &  -2.5&  -0.4&  -0.5&  -0.5&  -0.5&  -0.2&  -0.2&  -0.15&  -0.15& -0.15\\
    \texttt{sharp\_hi\_lim} &  0.3&  0.4&  0.5&  0.5&  0.5&  0.5&  0.5&  0.5&  0.5& 0.5\\
    \texttt{e\_mag\_lim    } &  10.0  & 10.0&  10.0&  10.0&  10.0&  10.0&  10.0&  10.0&  10.0& 10.0\\
    \texttt{snr\_lim    } &  5  & 5&  5&  5&  5&  5&  5&  5&  5& 5\\
    \texttt{OT\_lim    } &  8  & 8&  8&  8&  8&  8&  8&  8&  8& 8\\
    \texttt{flag\_lim}      &     8&  3&  8&  8&  3&  8&  8&  3&  3& 3\\
    \texttt{chi\_lo\_lim}   &     0&  0&  0&  0&  0&  0&  0&  0&  0& 0\\
    \texttt{chi\_hi\_lim}   &     2.0&  20.0&  20.0&  10.0&  20.0&  10.0&  10.0&  20.0&  20.0& 20.0\\
    \hline
    \end{tabular}
    \caption{Table of Dolphot parameters used to clean each single-band PSF photometry catalog before band matching. }
    \label{tab:cleaning}
    
\end{table}


\bibliography{n159.bib}{}
\bibliographystyle{aasjournalv7}



\end{document}